\documentclass[letterpaper,12pt]{article}
\usepackage[utf8]{inputenc}

\pdfoutput=1

\usepackage{graphicx}
\usepackage{natbib}
\usepackage{mathabx}

\newcommand\changed[1]{{#1}}

\title{Habitability of waterworlds: runaway greenhouses, atmospheric expansion and multiple climate states of pure water atmospheres}
\author{Colin Goldblatt}

\begin{document}

    \maketitle
    \begin{abstract}
There are four different stable climate states for pure water atmospheres, as might exist on so-called ``waterworlds''. I map these as a function of solar constant for planets ranging in size from Mars size to 10 Earth-mass. The states are: globally ice covered ($T_s\precapprox 245$\,K), cold and damp ($270\precapprox  T_s\precapprox 290$\,K), hot and moist ($350\precapprox T_s\precapprox 550$\,K) and very hot and dry ($T_s\succapprox 900$\,K). No stable climate exists for $290\precapprox  T_s\precapprox 350$\,K or $550\precapprox  T_s\precapprox 900$\,K. The union of hot moist and cold damp climates describe the liquid water habitable zone, the width and location of which depends on planet mass. At each solar constant, two or three different climate states are stable. This is a consequence of strong non-linearities in both thermal emission and the net absorption of sunlight. 

Across the range of planet sizes, I account for the atmospheres expanding to high altitudes as they warm. The emitting and absorbing surfaces (optical depth of unity) move to high altitude, making their area larger than the planet surface, so more thermal radiation is emitted and more sunlight absorbed (the former dominates). The atmospheres of small planets expand more due to weaker gravity: the effective runaway greenhouse threshold is about 35\,W\,m$^{-2}$ higher for Mars, 10\,W\,m$^{-2}$ higher for Earth or Venus but only a few W\,m$^{-2}$ higher for a 10 Earth-mass planet. There is an underlying (expansion neglected) trend of increasing runaway greenhouse threshold with planetary size (40\,W\,m$^{-2}$ higher for a 10 Earth-mass planet than for Mars). Summing these opposing trends means that Venus-size (or slightly smaller) planets are most susceptible to a runaway greenhouse.

The habitable zone for pure water atmospheres is very narrow, with an insolation range of 0.07 times the solar constant. A wider habitable zone requires background gas and greenhouse gas; N$_2$ and CO$_2$ on Earth, which are biologically controlled. Thus, habitability depends on inhabitance.
   \end{abstract}

\section{Introduction}

In this paper, I calculate the effect of changes to the vertical extent of pure water atmospheres on the radiation balance of waterworlds. As water atmospheres become hot and therefore thick, the area of the emitting and absorbing surface increases, modifying the planetary energy balance. This effect has been neglected in previous numerical calculations of the runaway greenhouse \citep{pollack-71,watson-ea-84,abe-matsui-88,k-88,Goldblatt2013}. With new calculations of the energy budget, I map out the various steady state climates of water worlds as a function of planetary mass (from Mars-size to 10 Earth masses) and incoming sunlight and hence describe the habitable zone. 

\changed{I consider pure water atmospheres only. For an atmosphere with $\sim 1$\,bar of non-condensible gas, this is gives a good approximation of the atmospheric structure when the atmosphere is hot, as the saturation vapour pressure of water would greatly exceed the dry air pressure. It is clearly a bad assumption for an atmosphere with $\gtrsim 10$\,bar of non-condensible gas, like that of present Venus---but then Venus has lost its water already.

For waterworlds, a pure water atmosphere is the baseline assumption at all temperatures. Waterworlds are a hypothetical class of planet with some orders of magnitude more water than Earth, such that there is a deep liquid ocean separated from the rocky core by a thick layer of water ice \citep{Leger2004}. These seem straightforward to form dynamically \citep{Raymond2004} and the mass-radius distribution of exoplanets between 1.5 and 4 Earth radii implies a hight frequency volatile rich planets \citep{WeissMarcy2014}. The isolation of the ocean-atmosphere from the rocky core gives rise to the assumption of pure atmospheres. To have a pure atmosphere given the existence of surface inventories of C or N, one would suppose that they were dissolved in the ocean. Carbon dissolves easily, as seen on Earth. Nitrogen would have to be dissolved in the form of ammonia or nitrate, which would be plausible in the absence of life. In our solar system, Europa and Ganymede are akin to waterworlds, and lack substantive atmospheres.}

In understanding planetary energy balance, those of us whose atmospheric intuition is based on Earth will naturally think that the amount of radiation which escapes to space should depend upon the surface temperature. Whilst much of the thermal spectrum is optically thick due to absorption by water, the region around 10\,$\mu$m is not, so a fraction of surface radiation does escape directly to space. Now, if the planet warms such that the ocean evaporates and the physical thickness of the atmosphere increases, the atmosphere will likewise become optically thick across the thermal spectrum such that no radiation from the surface will escape to space. Thermal emission would come dominantly from the atmospheric level where thermal optical depth is unity, which can be arbitrarily high above the surface but tends towards a constant temperature. For this reason, the thermal flux asymptotes to a constant with increasing surface temperature\citep[see][for reviews]{nakajima-ea-92,Goldblatt2012a}. Up-to-date calculations put this limit, which I refer to as the Simpson--Nakajima limit, around 280\,W\,m${^2}$ \citep{Goldblatt2013}. Likewise, the absorption of solar radiation asymptotes to a limit too \citep{k-88}, as both the levels of absorption and scattering optical depth of unity move high above the surface \citep{Goldblatt2013}.

Noting that emission of thermal radiation and absorption of sunlight happen in layers which are thin relative to the atmosphere facilitates straightforward scaling of plane-parallel radiative transfer calculations to an atmosphere which has become thick. I describe the methods for doing this briefly and discuss the changes in thermal flux emitted and solar flux absorbed in more detail. I then describe the four separate climate states of water worlds, \changed{including a hot and moist ($350\precapprox T_s \precapprox 550$\,K) stable state,} a finding which I believe to be novel. 

Note that I use the phrase ``atmospheric expansion'' as a shorthand to refer to increases in the vertical extent of the atmosphere relative to Earth's atmosphere as surface temperature increases, hence expansion of area of the emitting and absorbing levels (surfaces) within the atmosphere. It does not imply vertical velocity or acceleration; the atmospheres that I consider are all in hydrostatic balance.

\section{Methods}

\subsection{Model atmosphere and radiative transfer}

My model atmosphere and radiative transfer methods follow \citet{Goldblatt2013} directly. Hence I give a brief description only here; refer to the Methods and Supplementary Information of \citet{Goldblatt2013} for more detail. 

As I consider pure water atmospheres only, the moist adiabat which typically describes the structure of planetary atmospheres simplifies to the saturation vapour pressure curve of water. 

The mass range of the planets considered is from Mars \changed{(0.12 Earth mass)} to 10 Earth mass, via a 0.3 Earth mass planet, Venus \changed{(0.82 Earth mass)}, Earth and a 3 Earth mass planet. For the imaginary planets, \changed{the mass-radius relationship is approximated by} $M/M_\textrm{Eath} = (r/r_\textrm{Eath})^{2.7}$ \citep{Valencia2006}.

Surface temperatures range between 220\,K and 2000\,K. I define the structure as a the saturation vapour pressure curve when the surface temperature is less than the critical point, or a dry adiabat merging to the saturation vapour pressure curve when surface temperature is supercritical. I take the total water inventory to be that which gives a surface pressure of 260\,bar when the entire ocean evaporates, as on Earth. This is a quite arbitrary assumption, as the initial water inventory of a planet likely depends more on where it formed than its mass (compare Earth to Europa), and water inventory at some point in history depends on atmospheric evolution (compare Earth to Venus). However, given that the limiting fluxes are approached with less than a bar of atmospheric water \citep{Goldblatt2013}, this assumption is likely unimportant.

I omit clouds, but use an artificially high surface albedo (0.25) in lieu of cloud albedo. This assumption is frequently used in habitable zone, early Earth climate and runaway greenhouse studies \citep[e.g.][]{kpc-84,Goldblatt2009a}. This is not to say it is a good assumption \citep{gz-11}, only that the alternatives might be worse in this case. I have no intuition for cloud behaviour or feedbacks in these atmospheres, so sidestepping this allows me to focus attention on the clear sky behaviour, which is complex and fascinating enough itself. 

Radiative transfer calculations are done with the SMART code, written by David Crisp \citep{mc-96}, modified to include Rayleigh scattering by water \citep{Goldblatt2013}. Spectral data is taken from HITEMP2012 \citep{hitemp10} (HITRAN does not have sufficient lines for thick atmospheres). The sign conventions are positive downward (heats the planet)
for solar flux 
\begin{equation}
F_\mathrm{solar} = F_\mathrm{solar}^\downarrow - F_\mathrm{solar}^\uparrow
\end{equation}
and positive upward (cools the planet) for thermal fluxes 
\begin{equation}
F_\mathrm{thermal} = F_\mathrm{thermal}^\uparrow - F_\mathrm{thermal}^\downarrow
\end{equation}
and the net flux is positive downward
\begin{equation}
F  =  F_\mathrm{thermal} - F_\mathrm{solar}
\end{equation}.

\subsection{Scaling fluxes for expanded atmospheres}

A key property of runaway greenhouse atmospheres is that, for any given wavelength $\nu$, emission of thermal radiation and absorption of solar radiation both dominantly occur in layers which are thin relative to the whole atmosphere. I approximate this as occurring at a single level at which the optical depth is unity (altitude $z_{\tau=1}$). As $z_{\tau=1}$ becomes non-trivial compared to the planetary radius, $r_0$, the area of the emitting and absorbing surface of the planet increases. I thus define a scaling factor, 
\begin{equation}
s_\nu = \frac{r_{\tau=1}(\nu)}{r_0} = \frac{r_0 + z_{\tau=1}(\nu)}{r_0}
\end{equation}
to give the radiative flux absorbed or emitted per unit area of planetary surface from an expanded atmosphere. Changes to the size of the emitting and absorbing areas are not included in my SMART runs, so the flux from the expanded atmosphere is the product of the flux calculated in SMART and the scaling factor.

To find $z_{\tau=1}$, I begin with $p_{\tau=1}$ which SMART outputs as a function of wavelength. I solve for radius by integrating the hydrostatic equation up from the surface, allowing for variation in gravitational acceleration $g$ with radius ($g = g_0(r_0/r)^2$, where $g_0$ is gravity at the planetary surface).
\begin{equation}
\frac{dp}{dr} = -\rho g = -\rho g_0\left(\frac{r}{r_0}\right)^2 \label{e-hydrostatic}
\end{equation}
Given that the ideal gas law breaks down at the high temperatures considered here, I take density $\rho = \rho(p,T)$ from the IAPWS95 formulation \citep{wagner-pruss-02,junglas-08} and solve numerically to give $z = z(p)$. Vectors of altitude at which $\tau = 1$ and scaling factor $\mathbf{s}$, both as a function of wavelength, are found directly.

It is convenient to reduce $\mathbf{s}$ two to broadband scaling factors, for the solar radiation and thermal radiation. To do so, I take a weighted sum of the product of the vector scaling factor and the relevant irradiance. For thermal radiation, the spectrum of top of atmosphere outgoing thermal flux, $\mathbf{I^\uparrow_\mathrm{thermal,TOA}}$, is appropriate, giving
\begin{equation}
s_\mathrm{thermal} = \frac{\Sigma (\mathbf{s}\cdot\mathbf{I^\uparrow_\mathrm{thermal,TOA}})}{\Sigma\, \mathbf{I^\uparrow_\mathrm{thermal,TOA}}}
\end{equation}
For solar radiation, the spectrum of incoming sunlight at the top of the atmosphere, $\mathbf{I^\downarrow_\mathrm{solar,TOA}}$, is appropriate, giving
\begin{equation}
s_\mathrm{solar} = \frac{\Sigma (\mathbf{s}\cdot\mathbf{I^\downarrow_\mathrm{solar,TOA}})}{\Sigma\, \mathbf{I^\downarrow_\mathrm{solar,TOA}}}
\end{equation}
Fluxes from the expanded atmospheres, $F_\textrm{expand}$, are directly related to fluxes calculated by SMART, $F_\textrm{orig}$:
\begin{eqnarray}
F_\mathrm{expand,thermal} &=& s_\mathrm{thermal}F_\mathrm{orig,thermal} \\
F_\mathrm{expand,solar} &=& s_\mathrm{solar}F_\mathrm{orig,solar}
\end{eqnarray}
\changed{By derivation, $F_\textrm{expand}$ are fluxes \textit{per unit area of planetary surface}. These are useful because they enable one to consider changes in the energy budget with a fixed reference frame and so understand changes with increases in atmospheric size. However, they are somewhat non-physical in that a locally placed radiometer would measure the unexpanded flux.} 

Strictly, the plane parallel assumption employed in most radiation codes (SMART included) requires that the vertical extent of the atmosphere be smaller than the planetary radius. For the warmest atmospheres on Mars this becomes somewhat questionable, as optical depth of unity is around 1000\,km whereas the radius at the surface is 3390\,km. However, given that the limiting fluxes are determined in thin layers the effect of sphericity should be minimal and the scaling approach we use here should give an answer which is correct to first-order. The alternative, of full 3-D radiative transfer calculations, is prohibitively difficult and expensive. 

\subsection{Steady states for climate}

Overall temperature change of a planet is related to top of atmosphere fluxes---net sunlight absorbed by the planet and top of atmosphere thermal emission. For brevity:
\begin{eqnarray}
\mathcal{F}_\mathrm{solar} & = & F_\mathrm{solar,TOA} = F_\mathrm{solar,TOA}^\downarrow - F_\mathrm{solar,TOA}^\uparrow \\
\mathcal{F}_\mathrm{thermal} & = & F_\mathrm{thermal,TOA} = F_\mathrm{thermal,TOA}^\uparrow 
\end{eqnarray}
Scaling $\mathcal{F}_\mathrm{solar}$ by a constant $S/S_0$ can be used to reflect changing solar constant (due to orbital distance or stellar evolution), taking the reference value to be at Earth's orbit now, \changed{$S_0 = 1368$}\,W\,m$^{-2}$. Thus
\begin{equation}
\mathcal{F} = \mathcal{F}_\mathrm{thermal} - \frac{S}{S_0}\mathcal{F}_\mathrm{solar}  \label{e-ssclimate}
\end{equation}
$\mathcal{F}  =  0$ implies steady state for climate. With $\mathcal{F}$ defined positive upward $\frac{d\mathcal{F}}{dT_s}>0$ gives a stable steady state and $\frac{d\mathcal{F}}{dT_s}<0$ gives an unstable steady state. 

In my radiative transfer calculations, I assumed a fixed surface albedo, $\alpha_s = 0.25$. To include ice--albedo feedback in the climate calculations, I set surface albedo to represent an icy surface ($\alpha_s = 0.65$) when temperatures are low and vary smoothly to the standard value via a hyperbolic tangent \citep[e.g.][]{saltzman-02}:
\begin{equation}
\alpha_s = 0.45 - 0.2\tanh\left[\frac{\pi}{25}(T_s-255)\right]
\end{equation}
The purpose of this is to give a schematic representation of ice--albedo feedback, not to accurately predict the bifurcation points. 

A changed $\alpha_s$ changes the upward thermal flux vector $F_\mathrm{solar}^\uparrow$ only. $F_\mathrm{solar}^\uparrow$ is attenuated by both scattering and absorption. The source is, in general, a function of both scattering (of downward flux) and the surface source. Fortunately, the cold and thin atmospheres for which albedo is adjusted have negligible scattering optical depth, so the source is purely surface radiation and attenuation is only by absorption, so scaling $F_\mathrm{solar}^\uparrow$ to the new albedo can be done directly. 

\section{Results}

\subsection{The runaway greenhouse in expanded atmospheres}

\begin{figure}
\begin{center}
 \includegraphics{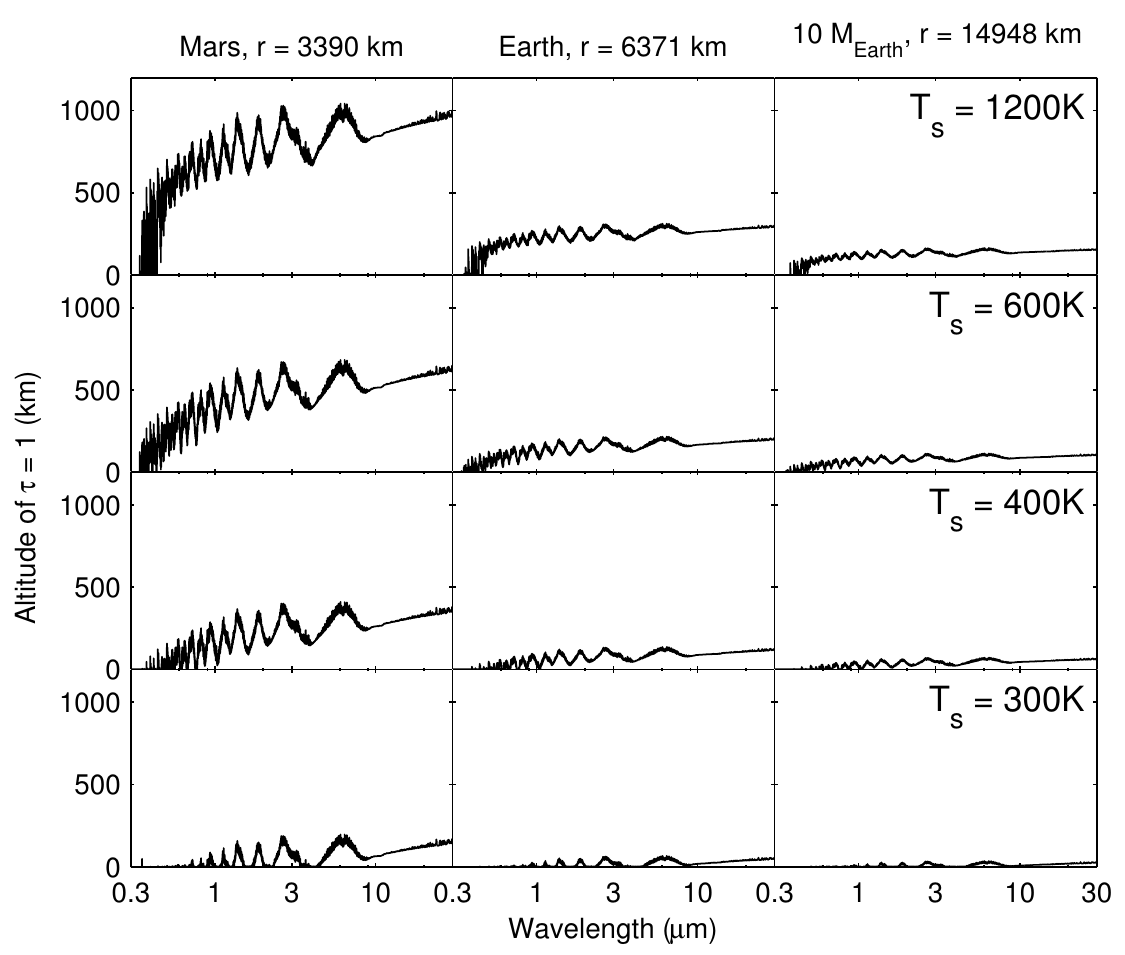}
\end{center}
\caption{Altitude at which absorption optical depth is unity ($z_{\tau=1}$) with changing temperature for three example planets of radius $r$. This is the height at which effective thermal emission to space and absorption of solar radiation will dominantly occur. Note how, for heavier planets $z_{\tau=1} \ll r$ so atmospheric expansion has little effect, whereas for small planets $z_{\tau=1} \sim r$ so atmospheric expansion is of first order importance. }\label{f-tau1example}
\end{figure}

The altitude at which absorption optical depth is unity is shown for Mars, Earth and a 10 Earth-mass planet (Figure \ref{f-tau1example}). For more massive planets, the vertical extent of the atmosphere is small compared to radius, even when the atmosphere is thick. By contrast, for small planets like Mars the atmosphere extends to high altitudes, especially when the surface is warm. In the middle, for Earth, optical depth of unity is at several hundred kilometres altitude for hot atmospheres, so expansion will matter. 

\begin{figure}
\begin{center}
 \includegraphics[width=\textwidth]{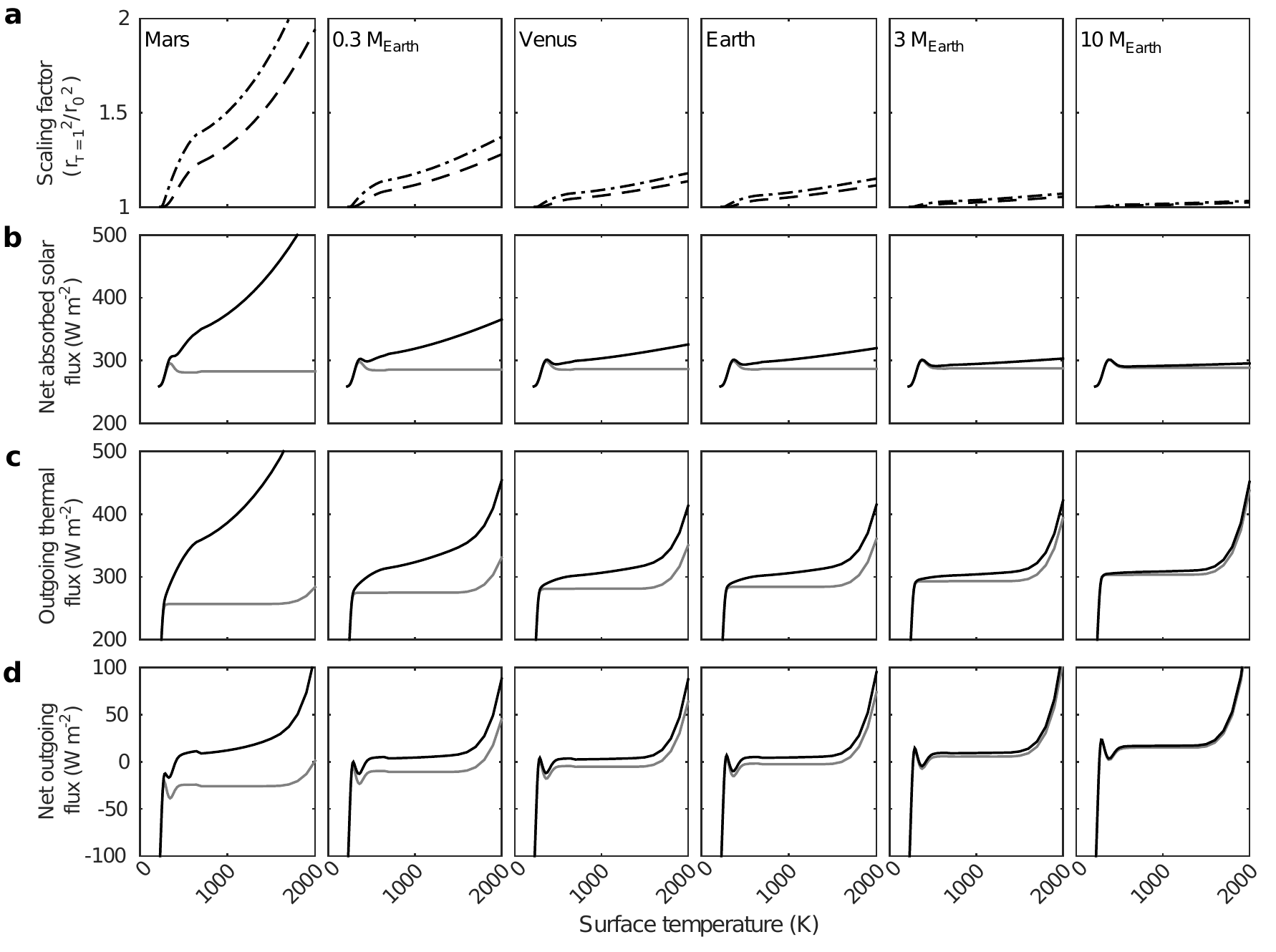}
\end{center}
\caption{For each planet considered (columns): ({\bf a}) Scaling factor; the ratio of squares of radius at which atmospheric optical depth is unity $r_{\tau=1} = z_{\tau=1}+r_0$ to the radius of the planet surface, $r_0$. This is the ratio of how much energy will be absorbed or emitted by the atmosphere relative an atmosphere with equivalent optical properties but negligible vertical extent. Solar is dashed, thermal dash-dot. ({\bf b}) Net absorbed solar flux (top of atmosphere incoming minus outgoing) given a solar constant equal to modern Earth's for atmospheres where expansion is neglected (grey) and considered (black). ({\bf c}) Outgoing thermal flux a the top of the atmosphere for atmospheres where expansion is neglected (grey) and considered (black). ({\bf d}) Net outgoing radiation (outgoing thermal minus absorbed solar) for atmospheres where expansion is neglected (grey) and considered (black). \changed{Negative values indicate planetary warming whereas positive values are planetary cooling}. Stable states have a zero net outgoing flux. }\label{f-expansion}
\end{figure}

Figure \ref{f-expansion} shows the scaling factor, absorbed solar flux ($\mathcal{F}_\mathrm{solar} $), outgoing thermal flux ($\mathcal{F}_\mathrm{thermal} $) and net flux ($\mathcal{F}$) for each planet as a function of temperature, both with expansion neglected and considered. Examine the expansion neglected fluxes first \changed{(grey lines)}, as these illustrate important points of the atmospheric physics. $\mathcal{F}_\mathrm{solar} $ initially increases with temperature, but subsequently decreases before asymptoting to a constant. Recall that increasing temperature is accompanied by increasing atmospheric mass, as ocean evaporates. The initial increase in $\mathcal{F}_\mathrm{solar} $ occurs as there is more molecular absorption as the atmosphere becomes thicker; referring back to Figure \ref{f-tau1example}, note how the atmosphere is optically thin for short wavelengths at low temperatures, but becomes optically thick \changed{at all wavelengths} as temperatures increase. With a thicker atmosphere still, the atmosphere becomes optically thick with respect to Rayleigh scattering too; bluer light which may otherwise have been absorbed at the surface, low \changed{in the} atmosphere or on the upward path is reflected by the atmosphere. This causes the decrease in $\mathcal{F}_\mathrm{solar} $. The asymptotic value for $\mathcal{F}_\mathrm{solar} $ reflects the balance between absorption and scattering cross sections \citep{Goldblatt2013}. 

Thermal fluxes are more straightforward, as in the infrared the atmosphere is absorbing but non-scattering. Increasing surface temperature increases surface emission by Stefan-Boltzman law. With low surface temperature the atmosphere is physically thin and optically thin. Surface radiation reaches space and $\mathcal{F}_\mathrm{thermal}$ increases with surface temperature. As temperature and atmospheric mass increase, the level from which thermal emission occurs moves up in the atmosphere, assumes a constant temperature and $\mathcal{F}_\mathrm{thermal}$ asymptotes to a constant \citep{nakajima-ea-92,Goldblatt2012a,Goldblatt2013}. When surface temperatures are $>1500$\,K, $\mathcal{F}_\mathrm{thermal}$ increases again as the high troposphere becomes sufficiently hot to emit in the 4\,$\mu$m water vapour window \citep{Goldblatt2013}. 

\changed{Still negleting atmospheric expansion}, planetary energy balance is determined $\mathcal{F} = \mathcal{F}_\mathrm{thermal} - \mathcal{F}_\mathrm{solar}$.
Combining cubic nonlinearities from both thermal and solar streams gives a quintic nonlinearity. The order of processes is: \changed{(1) Increasing surface temperature initially increases $\mathcal{F}_\mathrm{thermal}$, causing $\mathcal{F}$ to increase to a local maximum. (2) Increase in $\mathcal{F}_\mathrm{solar}$ due to absorption causes $\mathcal{F}$ to decrease to a local minimum. (3) Increased Rayleigh scattering decreases $\mathcal{F}_\mathrm{solar}$ allowing $\mathcal{F}$ to increase. (4) Both $\mathcal{F}_\mathrm{solar}$ and $\mathcal{F}_\mathrm{thermal}$ have reached asymptotic values, so $\mathcal{F}$ is constant.} (5) Finally, 4\,$\mu$m thermal emission increases $\mathcal{F}_\mathrm{thermal}$ and $\mathcal{F}$.

\begin{figure}
\begin{center}
 \includegraphics{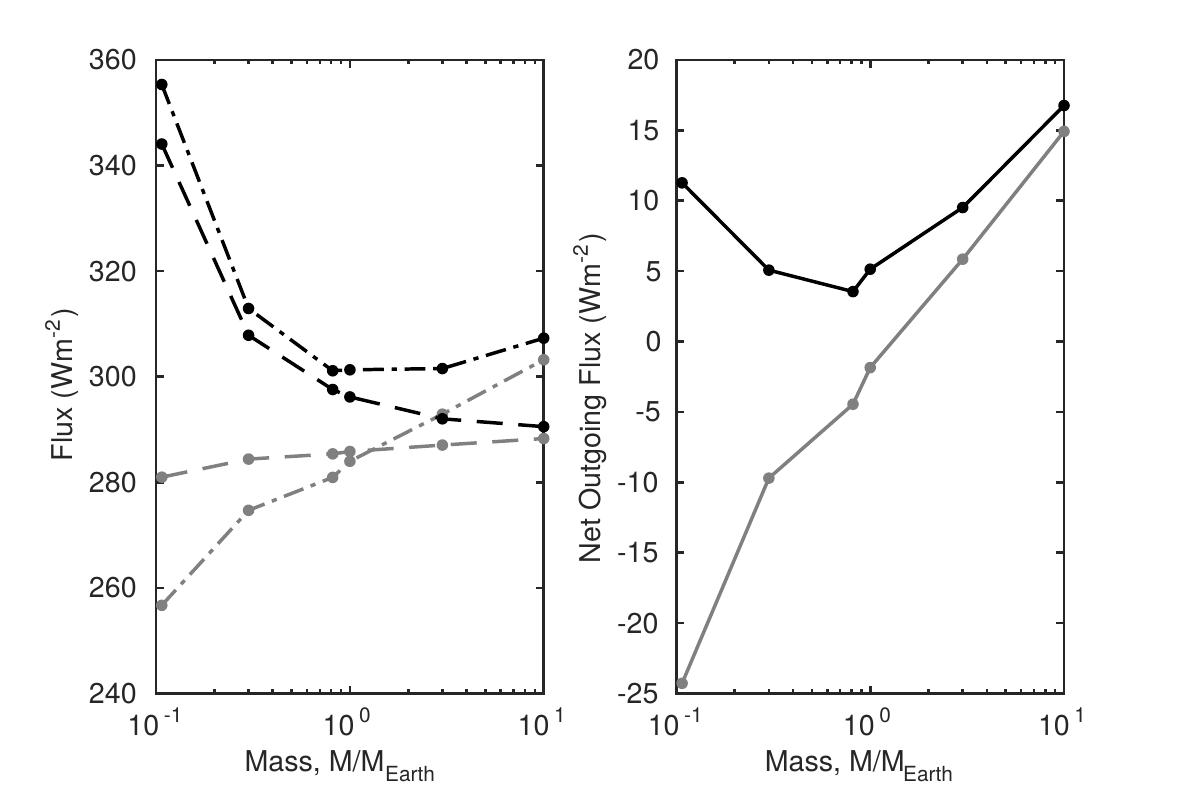}
\end{center}
\caption{({\bf a}) Radiation limits, taken as the flux at 645\,K, as a function of planetary mass. Net absorbed solar radiation as dashed line given a solar constant equal to modern Earth's, outgoing thermal radiation as dash-dot line, for atmospheres where expansion is neglected (grey) and considered (black). ({\bf d}) Net outgoing radiation (outgoing thermal minus absorbed solar) as a function of planet mass for atmospheres where expansion is neglected (grey) and considered (black).}\label{f-radiationlimits}
\end{figure}

Now consider how the asymptotic flux limits, still neglecting atmospheric expansion, vary with planetary mass (Figure \ref{f-radiationlimits}). Both $\mathcal{F}_\mathrm{solar}$ and $\mathcal{F}_\mathrm{thermal}$ are bigger on heavier planets, though somewhat different processes are in effect. Solar radiation absorbed depends on a balance between Rayleigh scattering and absorption cross sections. Rayleigh scattering cross section is independent of pressure, so scattering optical depth is proportional simply to the number of molecules in the optical path. By contrast, absorption cross section does depend on ambient pressure, as absorption lines are broadened by pressure. With a more massive planet (higher $g$), the same number of molecules will have a greater weight so exert more pressure, increasing the absorption cross section. Hence the ratio of absorption to scattering cross sections is larger, and more radiation is absorbed. 

Outgoing thermal flux depends on the temperature at the level of optical depth of unity. With atmospheric structure following the saturation vapour pressure curve, temperature is uniquely a function of pressure. With higher $g$, a smaller mass is required above some level to give the same pressure. Conversely, if to first approximation a fixed mass is required to give optical depth of unity, then under higher $g$ the base of this layer will experience higher pressure and hence be at higher temperature, so more radiation will be omitted. Due to pressure broadening of absorption lines, less mass is actually required at higher $g$, weakening this dependence \citep{pierrehumbert-10,Goldblatt2013}. This behaviour can be seen in Figure \ref{f-radiationlimits}.

Now consider atmospheric expansion. I have restricted my analysis to atmospheres in which hydrostatic balance applies (that is, not considered Europa or Ganymede, where massive hydrodynamic escape is confounding). The isothermal case is instructive; pressure decreases exponentially with altitude 
\begin{equation}
p = p_0\exp\left(\frac{-z}{H}\right)
\end{equation}
where the scale height $H =  \frac{RT}{Mg}$ (R is universal gas constant, M molar mass). Hence the altitude of some pressure level
\begin{equation}
z = -\frac{1}{g}\cdot\frac{RT}{M}\ln\left(\frac{p}{p_0}\right)
\end{equation}
is inversely proportional to gravity. Expansion thus affects small planets greatly, but big planets far less; the difference between expanded and unexpanded fluxes becomes increasingly large as planets become smaller than Venus (Figure \ref{f-radiationlimits}).  With water absorbing thermal radiation more strongly than solar radiation, the scaling factor for thermal radiation is always larger. The emitting surface becomes bigger, but this is largely offset by a bigger absorbing area (Figure \ref{f-expansion}). For planets of $0.3M_\textrm{Earth}$ and larger, these cancel such that $\mathcal{F}$ is still roughly constant between 500 and 1500\,K, but for Mars $\mathcal{F}$ increases with surface temperature through out this range.

Fluxes from the expanded atmosphere are a product of the unexpanded flux and the scaling factor. These have opposite trends with $g$. The consequence is a minimum in outgoing flux (i.e. maximum in susceptibility to a runaway greenhouse) for Venus-size or slightly smaller planets. The radiation limit considering expansion is around 10\,W\,m$^{-2}$ higher for Earth and Venus considering expansion. For a Mars-size planet, expansion puts the radiation limit 35\,W\,m$^{-2}$ higher. To put this in context, there has been a 30\% increase in the Sun's output through its main system lifetime which corresponds to a 70\,W\,m$^{-2}$ increase in absorbed solar flux for a planet at Earth's orbit, assuming a planetary albedo of 0.3. 

\subsection{Climate states and the habitable zone for pure water atmospheres}

\begin{figure}
\begin{center}
 \includegraphics{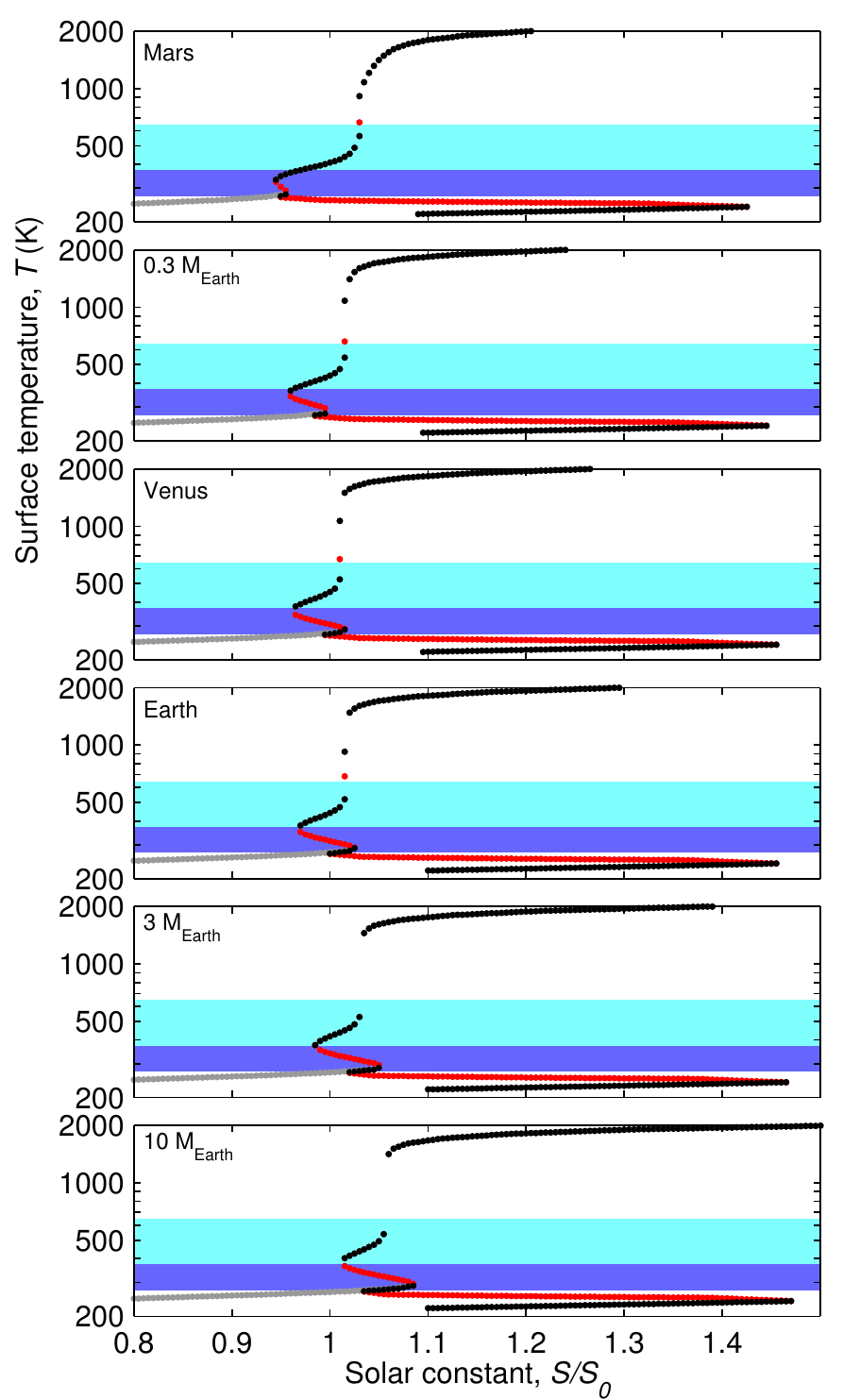}
\end{center}
\caption{Stable states of temperature (note log scale) for each planet as a function of solar constant. Black markers are stable steady states, red markers are unstable steady states. Grey markers are stable steady states when ice-albedo feedback has been omitted. Note that the lowest stable state would extend to arbitrarily low temperature, but my calculations cease at 220\,K. Blue shaded areas are the liquid water region (273\,K to 647\,K) with darker area being below 373\,K.}\label{f-tempandescape}
\end{figure}

Comparing the outgoing thermal flux to the net solar flux scaled by an arbitrary solar constant, allowing for expansion in both cases, allows us to map the stable states for water world climates (Figure \ref{f-tempandescape}). In addition to the quintic non-linearity which exists in $\mathcal{F}$, arising directly from the radiative transfer through the clear sky atmosphere, there is a cubic non-linearity introduced by ice-albedo feedback. The resulting function has a seventh degree non-linearlity. Hence we see four stable steady states, separated by three unstable steady states. 

Consider the four stable steady states. The coldest is global ice cover, with $T_s \precapprox 245$\,K, extending from arbitrarily low solar constant (not shown due to 220\,K minimum temperature) to $S/S_0 \approx 1.45$. Next is a cold and damp state ($270 \precapprox  T_s \precapprox 290$\,K) with liquid water. For both these states, stability arises from the negative temperature feedback with outgoing thermal radiation: the atmosphere is optically thin to thermal radiation, so increasing surface temperature gives increasing thermal emission to space. The two states are separated by the positive ice-albedo feedback. The cold and damp state becomes unstable when the negative temperature feedback breaks down---i.e. outgoing thermal radiation asymptotes to a constant. This bifurcation is typically thought of as the threshold for a runaway greenhouse \citep{nakajima-ea-92,Goldblatt2012a}, when absorbed solar flux exceeds the limit on thermal emission. However, a complete runaway does not occur here as there is an intermediate, hot and moist, stable state ($350\precapprox T_s \precapprox 550$\,K). The stability of this arises from a negative feedback between temperature and absorbed solar radiation: warming evaporates ocean and increases atmospheric mass, in turn increasing the amount of Rayleigh scattering such that less sunlight is absorbed. Only after the absorption of solar radiation asymptotes, when $T_s\succapprox 550$\,K, does the terminal runaway greenhouse occur. The final climate state is very hot and dry (typically $T_s \succapprox 900$\,K), above the critical point, so distinct liquid and vapour phases no longer exist. Stability arises from the re-establishment of a negative temperature feedback, with 4\,$\mu$m emission from the atmosphere increasing with temperature. 

\begin{figure}
\begin{center}
 \includegraphics{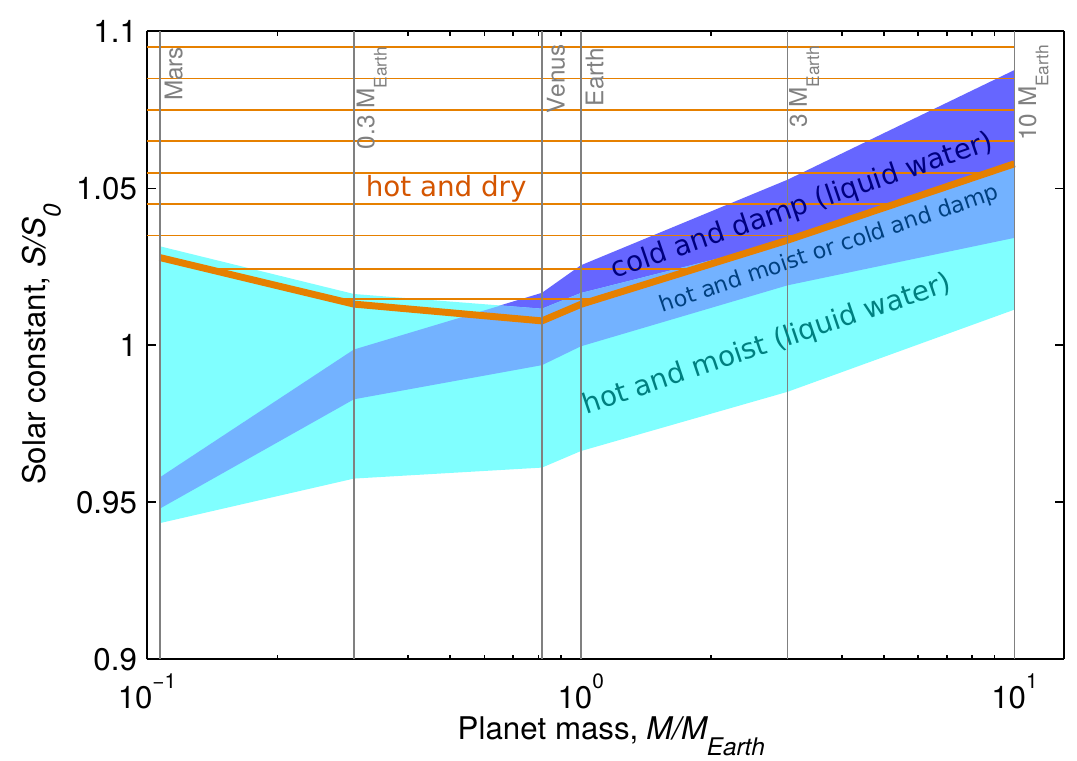}
\end{center}
\caption{Stable climate states mapped as a function of planet mass. Blue shaded regions have liquid water: darker area is the where cold and damp climate is stable, lighter area where hot and moist climate is stable and intermediate shading where both are stable. The orange hatched area bounded by a thick line is the stable region for very hot and dry climate. Completely ice covered regime is stable everywhere (not shown). }\label{f-liqwaterregion}
\end{figure}

Each stable steady state exists over some range in $S/S_0$, the position and width of which depends on planet size (Figure \ref{f-liqwaterregion}). The cold damp state exists at larger $S/S_0$ and has a larger range on bigger planets. Similarly, snowball-like conditions can exist at higher $S/S_0$ on bigger planets. The warm moist state has its base at higher $S/S_0$ on bigger planets, but its warm edge has a minimum around Venus-size. Likewise, the base of the hot dry state is minimal around Venus-size.

A counter-intuitive feature is that cold and damp climate is often stable \changed{only at relatively high} solar constants. This is a consequence of the existence of the ice albedo feedback. If this is neglected (grey dots in Figure \ref{f-tempandescape}), cold and damp conditions continue to arbitrarily low solar constant, making this state much broader than hot and moist conditions. Indeed, it is the hot moist state which is fundamentally narrow, depending on the Rayleigh scattering feedback which necessarily is limited to a small region of flux space. That this has its base at a lower solar constant than cold damp conditions does not have any profound significance.

\section{Discussion and Conclusions}

I have shown that there are four distinct stable steady states for climate in pure water atmospheres. Three arise directly from the pure sky radiative transfer, with the fourth contributed by ice-albedo feedback. Only one of these, cold damp climate with $270 \precapprox  T_s \precapprox 290$\,K would universally be considered habitable, though liquid water will also exist in the hot moist state with $350\precapprox T_s \precapprox 550$\,K. 

The existence of the hot moist stable state is, to the best of my knowledge, a novel finding. From a theoretical point of view, it is interesting because it owes its existence to a negative feedback whereby increasing planetary temperature increases the thickness of the atmosphere and the increase in Rayleigh scattering and therefore planetary albedo dominates. 

\changed{Practically, if hot moist climate turns out to be a robust result in less idealized atmospheres, it may be relevant to the evolution of Venus, the future of Earth, or similar planets. The hot water rich atmosphere would allow rapid hydrogen escape whilst the liquid ocean would mean that carbon could be sequestered in carbonates, not atmospheric CO$_2$. Forced by stellar evolution, the duration of this climate state would be hundreds of millions of years, which sufficient for loss of an Earth-ocean of water \citep{watson-ea-84}. Liquid water is the standard definition of habitability; hyperthermophiles could live at the surface and temperate conditions would exist at altitude. After sufficient water was lost, the planet would transition to a Dune-like climate state \citep{Herbert1965,abe-ea-11}, which is also habitable, and has a much higher runaway greenhouse threshold \citep{abe-ea-11}. This could add a billion years or more to the habitable period.}

I did not consider clouds, partly for expedience (many model runs would have been needed \citep[see][]{gz-11} which would have been very costly) and partly to keep the focus on fundamental results arising from clear-sky radiative transfer. However, one must asses whether these results would be robust to contact with clouds. For temperate atmospheres, wide variation of clouds alters the stable state temperature, but does not change the qualitative behaviour \citep{gz-11}. However, with more complicated qualitative behaviour here one can have little certainty: a wide parameter space for clouds should be considered, but this has to be left for the future. 

Accounting for expansion of the emitting and absorbing levels of runaway greenhouse atmospheres makes the threshold for a runaway greenhouse larger, especially for small planets, moving the inner edge of the habitable zone inward. Given the preferential affect on small planets, this changes the shape of the habitable zone with respect to planet mass.  

One of the more curious features of these results is the very narrow habitable zone for pure water atmospheres. In the sense of proof by contradiction, this demonstrates the importance of non-condensible gas to stable climate. Adding bulk gas (e.g. $\sim 1$\, bar N$_2$) would markedly weaken the non-linearity associated with increasing Rayleigh scattering as the atmosphere becomes thicker. It would also pressure broaden the absorption lines of water, so the cold and damp stable state would extend to lower solar constants. Adding much more still might extend the habitable zone to higher temperatures by scattering away sunlight. Adding greenhouse gas (e.g. CO$_2$), especially in the presence of a background gas \citep{glw-09}, can extend the habitable zone to lower temperatures. Our knowledge of the long term carbon and nitrogen cycles is somewhat rudimentary for Earth and non-existent for exotic planets.  Given that these gasses are biologically controlled, habitability depends on inhabitance and the width of the habitable zone is difficult to characterize.

\vspace{3cm}

\textbf{Acknowledgements} Thanks to Brendan Byrne and Kevin Zahnle for comments on the manuscript. This work was funded by C.G.'s NSERC Discovery grant and by NASA Planetary Atmospheres grant NNX11AC95G. 

\newpage
\bibliographystyle{plainnat} 
\bibliography{/home/czg/library/MendalyReflist/library}

\newpage

\newpage

\newpage

\newpage

\newpage


\end{document}